\documentstyle[12pt,aasms4]{article}
\begin{document}

\vspace*{0.5cm}

\title{Cluster Mass Estimate and a Cusp of the Mass Density
Distribution in Clusters of Galaxies}

\vspace{1cm}
\author{Nobuyoshi Makino$^{1,2}$ and Katsuaki Asano$^1$
}
\vspace{2cm}

\affil{\altaffilmark{1}Department of Physics, Ritsumeikan University\\
Kusatsu, Shiga 525-8577, Japan}
\affil{\altaffilmark{2}Department of Mechanical Engineering, Oita National 
College of Technology, 1666 Maki, Oita 870-0152, Japan}

\affil{\footnotesize e-mail: makino@oita-ct.ac.jp, sph10001@se.ritsumei.ac.jp}

\vspace{2cm}

\baselineskip=16pt

\abstract{ We study density cusps in the center of clusters of
galaxies to reconcile X-ray mass estimates with gravitational lensing
masses. For various mass density models with cusps we compute X-ray
surface brightness distribution, and fit them to observations to
measure the range of parameters in the density models. The Einstein
radii estimated from these density models are compared with Einstein
radii derived from the observed arcs for Abell 2163, Abell 2218, and
RX J1347.5-1145. The X-ray masses and lensing masses corresponding to
these Einstein radii are also compared. While steeper cusps give
smaller ratios of lensing mass to X-ray mass, the X-ray surface
brightnesses estimated from flatter cusps are better fits to the
observations. For Abell 2163 and Abell 2218, although the isothermal
sphere with a finite core cannot produce giant arc images, a density
model with a central cusp can produce a finite Einstein radius, which
is smaller than the observed radii.  We find that a total mass density
profile which declines as $\sim r^{-1.4}$ produces the largest radius
in models which are consistent with the X-ray surface brightness
profile. As the result, the extremely large ratio of the lensing mass
to the X-ray mass is improved from 2.2 to 1.4 for Abell 2163, and from
3 to 2.4 for Abell 2218. For RX J1347.5-1145, which is a cooling flow
cluster, we cannot reduce the mass discrepancy.}

\vspace{0.5cm} 

\noindent{\it Subject headings}: dark matter ---
galaxies:clusters;general --- gravitational lensing --- X-rays:
galaxies

\newpage

\section{Introduction}

\indent

Masses of clusters of galaxies has been estimated based on
observations of X-ray gas and by observation of gravitational lensing. 
Although cluster mass estimated by these methods should yield the
same values, they generally are are not, that is, the lensing mass is 
$2\sim3$ times larger than the X-ray mass(Grossman \& Narayan
1989). The X-ray mass reflects the gas distribution weakly bound by
the gravitational force. The predicted mass distribution has a large
core radius. Conversely, the lensing mass distribution is
strongly concentrated to produce distorted images around the center.

Many attempts have been made to reconcile the discrepancies between
the X-ray and lensing mass measurements have been made. Mass models
with ellipticity can reduce the mass required by the gravitational
lensing (Miralda-Escud\'{e} 1993). However, the decrease is not so
significant, especially, for Abell 1689 and Abell 2218
(Miralda-Escud\'{e} \& Babul 1995).  Miralda-Escud\'{e} \& Babul
(1995) presented a comparative analysis of lensing masses and X-ray
masses, and found that a temperature two times higher than the
observed value used be required to reconcile the discrepancy. Makino (1996)
investigated the temperature distribution of the gas in lensing
clusters, and showed that the gas temperature should significantly
increase toward the center. The pressure of the diffuse magnetic field
could support a gravitational force strong enough to produce giant
arcs (Loeb \& Mao 1994). However, this requirement for the magnetic
field leads to higher Faraday rotation measures in the lensing cluster
than in nearby clusters (Makino 1997). Moreover, Loeb \& Mao (1994)
proposed that in addition to the gas pressure, the bulk flow or
turbulent motions in the cluster structure would contribute to the
support of the X-ray gas. Gas dynamical simulations, however, cannot
produce strong dynamical motions (Evrard, Metzler, \& Navarro
1996).

Wu \& Fang (1996) pointed out, by comparing the masses for samples of
clusters derived from the X-ray to masses for samples from the
gravitational lensing clusters, that the mass discrepancy is more
significant around the cluster center than at large radii. Indeed, the
discrepancy between the X-ray mass and the weak lensing mass is
smaller than that between X-ray mass and the strong lensing mass which is
derived from giant arc images. Moreover, the discrepancy in cooling
flow clusters, which show strong peaks in the X-ray surface
brightness at their centers, can be resolved by taking account of the
effect of the peak in the surface brightness on the mass estimate
(Allen, Fabian, \& Kneib 1996; Allen 1997). This analysis is
not applicable to the non-cooling flow clusters.

The cluster density distribution has so far been believed to be smooth and
flat at the cluster center. Indeed, both isothermal spheres with
finite core radii $\rho\propto1/(r^2+r_{\rm c}^2)$ and modified
King models, $\rho \propto1/(r^2+r_{\rm c}^2)^{3/2}$ have been used to
estimate the cluster mass. However, using N-body simulations of
clusters of galaxies, Navarro, Frenk, \& White (1997) showed that
density cusps form at the center, and that the density profiles
have the same shape, independent of halo mass and of initial density
fluctuation spectrum. According to them, the central density profile
is well fit by an analytic form, $\rho\propto1/[r(r+r_{\rm s})^2]$
(hereafter referred to as the NFW profile). The NFW profile is found in
N-body/hydrodynamic simulations (Navarro, Frenk, \& White 1995;
Eke, Navarro, \& Frenk 1997). Furthermore, Carlberg et al. (1997)
found from measurements of galaxy velocity dispersions in clusters
that the NFW profile can reproduce the measured velocity
dispersion. At smaller scales, Hernquist (1990) derived an
analytically approximated density profile, $\rho\propto1/[r(r+r_{\rm
s})^3]$ from the de Vaucouleurs law for the surface brightness of
ellipticals. The Hernquist profile is in good agreement with numerical
simulations (Dubinski \& Carlberg 1991).

However, the value of the power of the density cusp at the center is
controversial because of the affect of gravitational softening and
poor resolution at small scales. Fukushige \& Makino (1997) showed
from simulations with particle numbers one order of magnitude higher,
that typical central density profiles are shallower than $\rho\sim
r^{-2}$ but steeper than $\rho\sim r^{-1}$. Moore et al. (1997)
also derived steeper inner density profiles, $\rho\sim r^{-1.4}$
from cosmological N-body simulation with higher resolution.
Furthermore, Evans \& Collett (1997) found a steady-state,
self-consistent, cusped solution of the collisional Boltzmann equation
corresponding to $\rho\sim r^{-4/3}$. Their solution is not
thermodynamically stable but is dynamically stable.

In this paper, we report on a study of the effect of density cusps
in clusters of galaxies on the lensing mass and the X-ray mass. 
Assuming simple cluster mass models with cusps, i.e. spherically
symmetric clusters with isothermal gas, we explicitly show that it is
possible that the density cusp at the cluster center reduces the mass
discrepancy between the X-ray mass and the lensing mass.
In \S 2 we show our two step method: first we fix
the parameters in the mass density models by comparing the estimated
X-ray surface brightness profiles with the observed ones, and second,
derive the estimated Einstein radii in the models and compare them
with the observed Einstein radii from arcs.  In \S 3, our method is
applied to the observation of the lensing clusters, Abell 2163, Abell
2218 and RX J1347.5-1145, the X-ray masses and lensing masses are
derived and the mass discrepancy are examined. In \S 4 we summarize
our conclusions, and discuss our results.

Throughout this paper, we assume $H_0=50{\rm
km\,s^{-1}\, Mpc^{-1}}$, $\Omega_0=1$ and $\Lambda_0=0$.

\section{Method of analysis}

We consider a spherically symmetric cluster of galaxies with an X-ray
emitting gas cloud, which is massive and distant enough to produce
giant arcs around the center. We assume that the gas cloud of
temperature $T$ is isothermal and in hydrostatic equilibrium, and that
the total mass density profile $\rho(r)$ in the cluster can be
described by a cusped density profile,
\begin{equation}
 \rho(r)=\frac{\rho_0}{x^{\mu} (1+x^{\nu})^{\lambda}},
\label{eq:rhodef}
\end{equation}
where $\rho_0$ is the normalization parameter and $x=r/r_{\rm s}$;
$r_{\rm s}$ is the scaling parameter. For example, the mass density
with $(\mu,\nu,\lambda)=$ (0,2,1) is an isothermal sphere
with a finite core radius and (1,1,2) corresponds to the NFW profile.
N-body simulations show that mass densities at the cluster scale decline as
$r^{-3}$ at large radii (Navarro, Frenk, \& White 1995; Navarro,
Frenk, \& White 1996). We adopt the parameters that satisfy
 $\mu+\nu \lambda=3$, except for an isothermal sphere.

The gas density distribution, $\rho_{\rm g}$, satisfies the equation of
hydrostatic equilibrium:
\begin{equation}
  \frac{k T}{\mu_{\rm m} m_{\rm p}}\frac{d \ln{\rho_{\rm g}}}{d r}
  =-\frac{G M(r)}{r^2},
\label{eq:hydro}
\end{equation}
where $\mu_{\rm m}$ and $m_{\rm p}$ denote the mean molecular weight
(we adopt $\mu_{\rm m}=0.59$ below) and the proton mass, respectively.
Here we have neglected the contributions of gas and galaxies
to the total mass $M(r)$ within a radius $r$. The total mass $M(r)$
is estimated by integrating equation (\ref{eq:rhodef}), and the gas
density profile is uniquely determined from equation (\ref{eq:hydro}), if
the gas temperature is given by observation. The gas density is
expressed by two parameters, $r_{\rm s}$ and $b$, defined as 
\begin{equation}
  b=\frac{8 \pi G \mu_{\rm m} m_{\rm p} \rho_0 r_{\rm s}^2}{27 k T}.
\label{eq:bdef}
\end{equation}
For the NFW profile ($\mu=1$, $\nu=1$, $\lambda=2$)
one finds that the gas density profile has the analytic form:
\begin{equation}
  \rho_{\rm g}(r)=\rho_{\rm g0} \exp{\left[-\frac{27}{2} b 
\left(1-\frac{\ln{(1+x)}}{x}\right)\right]},
\end{equation}
where $\rho_{\rm g0}$ is the central gas density (Makino, Sasaki, \&
Suto 1997).

To compare directly our gas density models with observation, we
compute the X-ray surface brightness given by
\begin{equation}
  I(\theta) \propto \int_{D_{\rm l} \theta}^{r_{\rm cut}}
\frac{\rho_{\rm g}^2}{\sqrt{r^2-(D_{\rm l} \theta)^2}} r dr,
\label{eq:bright}
\end{equation}
where $D_{\rm l}$ is the angular diameter distance of the cluster from
the observer. We have introduced a finite cut-off radius $r_{\rm cut}$
to avoid divergence in equation (\ref{eq:bright}). Since a mass
density profile which falls off as $r^{-3}$ at large radii cannot bind the
isothermal gas, the gas density is flat at large radii. The flat gas
density at infinity leads to divergence in the surface brightness.

A simpler mass density profile has been extensively used to obtain the 
cluster mass from X-ray observations (e.g., Henry, Briel, \&
Nulsen 1993; Elbaz, Arnaud, \& B\"{o}hringer 1995). The observed X-ray
emission is conventionally fitted with an isothermal $\beta$-model
(see, Sarazin 1988)
\begin{equation}
  I(\theta)=I_0 \left[ 1+\left(
  \frac{\theta}{\theta_{\rm c}}\right)^2 \right]^{-3\beta+\frac{1}{2}},
\label{eq:beta}
\end{equation}
where $I_0$ is the central surface brightness, $\theta_{\rm c}$ is the
angular core radius of the X-ray surface brightness, and $\beta$ is
the slope parameter. In the $\beta$-model, the gas density given by
from equation (\ref{eq:beta}) with $r_{\rm cut}=\infty$ is:
\begin{equation}
  \rho_{\rm g}(r)=\rho_{\rm g0} 
  \left[ 1+\left(\frac{r}{r_{\rm c}}\right)^2 \right]^{-3\beta/2},
\label{eq:rhogbeta}
\end{equation}
where $r_{\rm c}=\theta_{\rm c} D_{\rm l}$. Substituting equation
(\ref{eq:rhogbeta}) into equation (\ref{eq:hydro}), one obtains the total mass.
Furthermore, differentiating equation (\ref{eq:hydro}) with respect to
$r$, one finds the total mass density profile
\begin{equation}
  \rho(r)=\frac{3 \beta k T}{4 \pi G \mu_{\rm m} m_{\rm p} r_{\rm c}^2} 
 \frac{3+(r/r_{\rm c})^2}{\left[ 1+(r/r_{\rm c})^2 \right]^2}.
\label{eq:rhobeta}
\end{equation}
This density distribution is different from the cusped density
distribution in equation (\ref{eq:rhodef}). The mass density in
equation (\ref{eq:rhobeta}) is non-singular at the center. The cluster
with a mass density given by (\ref{eq:rhobeta}) would not have a large
enough mass to produce giant arcs around the center.

On the other hand, we know that clusters do produce giant arcs. This
imposes constraints on possible mass distributions, independent of
X-ray mass estimates. In spherically gravitational lensing, if a lens
is on the line of sight to a background source, the source will appear
as arc images around a radius $\theta_{\rm E}$ at which the flux of a
lensed image is apparently divergent in the limit of the geometrical
optics. We refer to the radius $\theta_{\rm E}$ as the Einstein
radius. In observations of gravitational lensing we assume that the
Einstein radius, $\theta_{\rm E}$, is equal to the distance from the
lens center to the giant arcs.

Now for a cusped density profile we determine the parameters $b$ and
$\theta_{\rm s}$ in equation (\ref{eq:rhodef}) and (\ref{eq:bdef}) on
the basis of X-ray data on the surface brightness, and we estimate the
cluster mass and the X-ray Einstein radius, $\theta_{\rm E,X}$ as
follows. The X-ray estimated Einstein radius is independent of the
observed radius $\theta_{\rm E}$, and the difference between the two
reflects the mass discrepancy. In the case of a spherically symmetric
lens, the Einstein radius is completely described by the surface
mass density $\Sigma(\theta)$ defined by
\begin{equation}
  \Sigma(\theta)=2 \int_{D_{\rm l} \theta}^{\infty} 
  \frac{\rho(r)}{\sqrt{r^2-(D_{\rm l} \theta)^2}}r dr.
\label{eq:sigdef}
\end{equation}
and the critical surface mass density
\begin{equation}
 \Sigma_{\rm cr}=\frac{c^2}{4 \pi G} \frac{D_{\rm s}}{D_{\rm l} D_{\rm ls}},
\end{equation}
where $D_{\rm s}$ and $D_{\rm ls}$ are the angular diameter distances
from the observer to the source and from the lens to the source,
respectively. We define the projected mass within $\theta$ by the
parameters $b$ and $r_{\rm s}$ determined from X-ray observation
\begin{equation}
  m(\theta)=2 \pi \int_0^\theta \Sigma(\theta')
\theta' d\theta'.
\label{eq:promass}
\end{equation}
The X-ray estimated Einstein radius should satisfy
\begin{equation}
  m(\theta_{\rm E,X})=\pi \theta_{\rm E,X}^2 D_{\rm l}^2\Sigma_{\rm cr},
\label{eq:E.R.}
\end{equation}
where we use $\rho(r)$ with the parameters determined from the X-ray
observation. 

We define the projected lensing mass within the measured Einstein
radius using the observed quantities, $\Sigma_{\rm cr}$, $\theta_{\rm
E}$, and $D_{\rm l}$, as follows:
\begin{equation}
  m_{\rm lens}(\theta_{\rm E})\equiv\pi \theta_{\rm E}^2 D_{\rm l}^2\Sigma_{\rm cr}.
\end{equation}
Note that the definition of the projected lensing mass is applicable
to the region $\theta_{\rm E}$ because the giant arcs reflect the
mass distribution within the measured Einstein radius only. Thus, the
projected X-ray mass compared to the projected lensing mass is only 
constrained by the region within the measured Einstein radius
$\theta_{\rm E}$. 

We can estimate an Einstein radius from the density models. We define
the projected X-ray mass by the mass surface density $\Sigma$,
determined from the X-ray observation:
\begin{equation}
  m_X(\theta_{\rm E})=2\pi\int^{\theta_{\rm E}}_0
  \Sigma(\theta^\prime)\theta^\prime d\theta^\prime,
\end{equation}
where we again use $\rho(r)$ with the above parameters determined by
the X-ray observations. This $m_{\rm X}(\theta_{\rm E})$ is the X-ray
mass to be compared with the lensing mass.

For the NFW profile, we can analytically derive $m(\theta_{\rm E})$
from the mass distribution. The detailed derivation of the Einstein
radius for the NFW profile was found in Bartelmann (1996) and Maoz et
al.  (1997). However, for most of our density models, the projected
mass cannot be analytically estimated from the mass distribution. We
numerically integrate the right hand of equation (\ref{eq:promass}) to
estimate the Einstein radius $\theta_{\rm E,X}$.

Our method is summarized as follows: we compute the X-ray surface
brightness for the various density mass models described by equations
(\ref{eq:rhodef}) and (\ref{eq:rhobeta}). The estimated surface
brightness is fitted to the X-ray observation of lensing clusters to
fix the parameters in the density models. Once the parameters are
determined, we can compare the measured Einstein radius with the
Einstein radius estimated from X-ray observation.  We also estimate the
projected mass and the Einstein radius for each density model.

\section{Results}

\indent

To study the effect of central mass concentration on the X-ray surface
brightness and gravitational lensing, we adopt the following
parameters in equation (\ref{eq:rhodef}): $\nu=1$ group;
$(\mu,\nu,\lambda)=$ (1,1,2), (1.2,1,1.8), (1.4,1,1.6), (1.6,1,1.4),
(1.8,1,1.2), and $\lambda=1$ group; (1,2,1), (1.2,1.8,1), (1.4,1.6,1),
(1.6,1.4,1), (1.8,1.2,1). In addition to the cusped mass density, we
compute the projected mass and the Einstein radius from the mass
density derived from the $\beta$-model in equation (\ref{eq:rhobeta}),
and from the isothermal sphere with a finite core radius for
comparison. The cut-off radius is fixed as $r_{\rm cut}=10 r_{\rm s}$
in our calculation of equation (\ref{eq:bright}). For the isothermal
model $(\mu,\nu,\lambda)=$ (0,2,1), we exceptionally fix $r_{\rm
cut}$ as $10^4 r_{\rm s}$. The gas densities derived from the cusped
density profiles are flat at radii $\geq 10 r_{\rm s}$. Since the
cusped density profiles decline as $r^{-3}$, the gas density cannot be
strongly bound. Unless the gas distribution at large radii declines
faster than $r^{-0.6}$ corresponding to $\beta=0.4$ in the
$\beta$-model, the gas density at the radius $r\sim D_{\rm l}\theta$
mainly contributes to the integral in equation (\ref{eq:bright}). As
will be shown below, the cut-off radius of $10 r_{\rm s}$ is larger
than the radius in which X-ray emissions from the clusters are
detected. Thus, the choice, $r_{\rm cut}=10 r_{\rm s}$, is reasonable
for our aim. The general behavior of the surface brightness estimated
from the cusped density profile for the $\nu=1$ group is discussed in
Suto, Sasaki, \& Makino (1998).

We apply our method of estimations to three clusters, Abell 2163,
Abell 2218 and RX J1347.5-1145. They are X-ray clusters in which giant
arcs have been found.  Their cluster redshifts, gas temperature and
observed Einstein radii are summarized in Table 1. We also list the
parameters of the $\beta$-model, $\beta$ and $\theta_{\rm c}$, derived
from our fitting results of the observed surface brightness in Table
1. 

\subsection{The X-ray surface brightness}

We illustrate the behavior of the surface brightness derived from 
cusped density profiles. In Figure 1 we show the X-ray surface
brightness estimated from the cusped profiles for $b=0.7$. The surface
brightness for density models with $\nu=1$ is plotted in Figure
1(a). The solid line indicates the surface brightness for
$(\mu,\nu,\lambda)=(1,1,2)$. The dotted line, short dotted line, long
dotted line and dashed dotted line show the surface brightness
profiles for $(\mu,\nu,\lambda)=(1.2,1,1.8)$, $(1.4,1,1.6)$,
$(1.6,1,1.4)$, and $(1.8,1,1.2)$, respectively

The surface brightness profiles with the parameters $\mu=1$, 1.2 and
1.4 are similar in shape. They show a flat core around the center and
a steep power law shape at large radii. Makino et al. (1998) showed
that the gas density for $\mu=1$ is in good agreement with the
$\beta$-model in equation (\ref{eq:rhogbeta}) which gives a good fit
to the surface brightness of real clusters of galaxies. Thus, the
surface brightness profiles for $\mu=1$, 1.2 and 1.4 are expected to
agree with the observation of clusters of galaxies. However, the
surface brightness for $\mu=1.8$ shows a single power law shape and no
finite core. It would be difficult to fit the surface brightness for
$\mu=1.8$ to observations. For $\mu=1.6$ the surface brightness is
marginal. The core around the center deviates slightly from a flat
profile. While its surface brightness could agree with cooling flow
clusters, it is not a good fit to clusters with a flat and large core
radius like the Coma cluster.

The decrease of the core radius for larger $\mu$ is due to the strong
concentration of the mass density at the center for large $\mu$.
Especially for $\mu=1.8$ the core radius is not found in Figure 1(a). 
The behavior of the surface brightness for $\mu=1.8$ is substantially
the same as $\mu=2$. The gas density in the density
model with $\mu=2$ is divergent at the center. In order to see the
behavior of gas density for $\mu=2$, we expand the integrand in
equation (\ref{eq:hydro}) around the center, $x\ll 1$:
\begin{equation}
 \label{eq:approx}
  -\frac{d\ln\rho_{\rm g}}{d\ln r}\approx\frac{4\pi G\rho_0\mu_{\rm m} m_{\rm p}}{kT}
  \frac{r_{\rm s}^{3-\mu}}{r}\int^{r/r_{\rm s}}_0 x^{2-\mu}(1-\lambda x^\nu)dx.
\end{equation}
If $\mu=2$, the right hand in equation (\ref{eq:approx}) is finite at
the limit $r\rightarrow 0$. The gradient of the gas density is finite
at the center.  The density model with $\mu=2$ cannot have a flat core
in surface brightness under the assumption of the isothermal gas.
For $\mu=1.6$, the X-ray core radius is extremely small.

In Figure 1(b) we plot the surface brightness for the parameters 
$(\mu,\nu,\lambda)=(1.2,1.8,1)$, $(1.4,1.6,1)$, $(1.6,1.4,1)$,
and $(1.8,1.2,1)$. The surface brightness in the $\lambda=1$ model
is steeper than in the $\nu=1$ model at large radii for the same value
of $b$. It is evident from equation (\ref{eq:approx}) that a cusped density
profile with a smaller $\nu$ and a larger $\lambda$ has a larger core
radius. For example, the surface brightness for $(\mu,\nu,\lambda)$=
(1,1,2) in Figure 1(a) is flatter than that for (1,2,1) in
Figure 1(b).

\subsection{Abell 2163}

Abell 2163 is a distant and elliptical cluster of galaxies. It is
classified as an Abell richness class 2 cluster.The redshift is
$z_{\rm l}=0.201$. Abell 2163 has a very high temperature of 12.4keV
(Holzapfel et al. 1997) and X-ray luminosity in the 2-10 keV band of
$L_{\rm X}=6.0 \times 10^{45}$ ergs ${\rm s^{-1}}$ (Elbaz et
al. 1995). Two arcs have been observed around the central galaxy. The
measured Einstein radius is $15.\!''6$, and its redshift is 0.728
(Miralda-Escud\'{e} \& Babul 1995).

We fit the surface brightness of the cusped density profiles to the
observations by the $\chi^2$ minimization, and determined the values of
$b$ and $\theta_{\rm s}$. In Figure 2, we plot the observed X-ray
surface brightness profiles of Abell 2163 derived from the {\it ROSAT}
HRI images and the best-fit profiles for various $(\mu,\nu,\lambda)$
and the $\beta$-model. 

The center of the surface brightness profile is located on the X-ray
maximum of the cluster emission. The errors are 1-$\sigma$ errors and
account for Poisson statistics. A flat exposure map is assumed.
That is, vignetting effects are not taken into account in the data.
The original data analysis can be found in Elbaz et al. (1995).

All the models are good fits at large radii, but the behavior of the
profile for each of our models is very different in the flat core region. 
The contribution of the density cusp to the surface brightness profile
is significant in the flat core region. Thus, we confine our
attention to the behavior of the best-fit profiles within $\theta_{\rm
c}$ below.

The best-fit profiles for the $(\mu,\nu,\lambda)=$ (0,2,1), (1,1,2)
and (1.2,1,1.8) models are very similar each other. They approximately
reproduce the core structure of the X-ray emission. These three
profiles deviate slightly upwards from the $\beta$-model within
$\theta_{\rm c}$ toward the center. The surface profile for
$(\mu,\nu,\lambda)=$ (1.4,1,1.6) model is steeper than the models with
$\mu\leq1.2$, but is within the error bars. The X-ray profiles of more
centrally concentrated models, $(\mu,\nu,\lambda)=$ (1.6,1,1.4) and
(1.8,1,1.2) are too steep in the central region to agree with the
innermost data point within the error bars. The best-fit profiles of
these models become steeper as $\mu$ increases.  This tendency is
also seen in Figure 2(c) for the $\lambda=1$ group.  The behaviors of
the profiles for the $\lambda=1$ group within $\theta_{\rm c}$ are
slightly flatter than ones for the $\nu=1$ group for the same value of
$\mu$. The deviation from the best-fit $\beta$-model for the
$\lambda=1$ group is relatively small.

From the reduced $\chi^2$ values in Table 2(a) and Figure 3(a), we
find that the $\beta$-model is the best fitting model for the X-ray
emission in Abell 2163, and that the cusped density profiles for
$\mu=1,\,1.2$ and 1.4 marginally agree with the data. It is apparent
from Figure 3(a) that the reduced $\chi^2$ values for the cusped
models change over the range $1.3\sim1.8$, except for $\mu=1.6$ and
$\mu=1.8$. Although we present the reduced $\chi^2$ only for the
$\nu=1$ model, the behavior of the reduced $\chi^2$ for the
$\lambda=1$ group is similar. The relatively large values of the
reduced $\chi^2$ are caused by substructures in Abell 2163 which
appear in the X-ray image (Elbaz et al. 1995). If we could get rid of
the contribution of the substructures, the models for $\mu \leq 1.4$
would be in agreement with the data.

We estimate the X-ray predicted Einstein radii $\theta_{\rm E,X}$ and
the projected masses $m_{\rm X}$ within the measured Einstein radius
$\theta_{\rm E}$ for Abell 2163. The projected lensing mass of Abell
2163 is estimated to be $4.13\times10^{13} M_\odot$. The results are
summarized in Table 2. For the $\nu=1$ model the mass ratio and
Einstein radius ratio are plotted in Figure 3 (b) and (c).
For the $\nu$=1 group, the ratios of the
projected lensing mass to the projected X-ray mass monotonically
approach one as $\mu$ increases, but none is less than one. 
Although a larger $\mu$ decreases the mass ratio $m_{\rm lens}/m_{\rm
X}$, the deviation of the estimated X-ray surface brightness around
the center from the data is significantly large. This tendency is also
found in the ratio $\theta_{\rm E,X}/\theta_{\rm E}$. The ratio
increases as $\mu$ increases for the $\nu=1$ group. Interestingly, a
finite Einstein radius can form in the cusped density profiles
although the giant arcs cannot be produced in the density models with
a finite core. This implies that the mass concentration around the
center is required to produce giant arcs. From the X-ray and the
lensing analysis, our best mass density model is
$(\mu,\nu,\lambda)$=(1.4,1,1.6) for Abell 2163. This density profile
increases toward the center as $\rho\propto r^{-1.4}$. The estimated
Einstein radius for the our best model amounts to 47\% of the observed
value. The projected mass for our best model is 57\% larger than the
mass estimated from the best-fit $\beta$-model.

For the $\lambda$=1 group, we obtain the same results as for the
$\nu=1$ group from Table 2(a). We, however, note that the change in
$b$ for $\nu$=1 groups is less sensitive to $\mu$ than in the
$\lambda$=1 groups. The mass discrepancy is not improved for the
$\lambda$=1 group as significantly as for the $\nu$=1 group.

\subsection{Abell 2218}

Abell 2218 is an Abell richness class 4 cluster at $z_{\rm l}$=0.175. The
radial velocity dispersion of cluster galaxies is very high,
1370km\,${\rm s^{-1}}$ (Le Borgne, Pello, \& Sanahuja
1992). About 30 arcs have been identified in this cluster. The
configuration of these arcs strongly suggests that this cluster has
two mass clumps (Kneib et al. 1995). While there is a secondary bright
clump around galaxy \#244 in a optical observation (Pell\'{o} et
al. 1992), the X-ray image shows only one maximum close
to the position of the cD galaxy. We consider one giant arc \#359 with
a radius of $20.\!''8$. The redshift of the source galaxy is 0.702
(Miralda-Escud\'{e} \& Babul 1995). This arc is mainly caused by the
central clump around the cD galaxy. Thus, we can neglect the second mass
clump. The X-ray luminosity, $L_{\rm X}$, is $0.7 \times 10^{45}$ ergs
${\rm s^{-1}}$ (0.1-2.4 keV), and the gas temperature is measured to
be 7.2keV (Mushotzky \& Loewenstein 1997).  Birkinshaw \& Hughes
(1994) reported $\theta_{\rm c}=60''$ and $\beta=0.65$. On the other
hand, Markevitch (1997) reported that the X-ray emission within a core
radius of $60''$ is resolved into several components. Markevitch
speculated that this may be caused by lensed X-ray emission, a merger
shock or a gas trail of an infalling subgroup, or both lensing and a
merger. Under these assumptions, he reanalyzed the X-ray profile and
obtained $\theta_{\rm c}=26''$ and $\beta=0.49$. We do not consider
this possibility. We simply fit the surface brightness data with our models
and we ignore the substructures in the cluster.

In Figure 4, we show the observed X-ray surface brightness profiles of
Abell 2218 derived from the {\it ROSAT} HRI images and the best-fit
profiles for various values of $(\mu,\nu,\lambda)$. The analysis of
the data for Abell 2218 is the same as for Abell 2163.  The detailed
description of the data analysis is found in Squires et al. (1996)

We also plot the best-fit $\beta$-model with $\beta=0.56$ and
$\theta_{\rm c}=53''$. The data points are smoothly distributed in
comparison with the data of Abell 2163. The behavior of the best-fit
profiles of the $\nu=1$ and $\lambda=1$ groups for Abell 2218 is
basically similar to those for Abell 2163. All the profiles are in
excellent agreement with the data in the power law region at large
radii.  All of the surface brightnesses estimated from the cusped models
are steeper than that from the best-fit $\beta$-model at the flat core
region. The best-fit profile becomes steeper as
$\mu$ increases. All of the profiles of the $\nu=1$ group deviate
from the error bars.  The surface brightness of the $\lambda=1$ group
pass inside the error bars only for $\mu \leq 1.2$.

We list results of the lensing analysis for Abell 2218 in Table 2(b).
For the $\nu=1$ model, the results are plotted in Figure 3.  The
projected lensing mass of Abell 2218 is $6.3\times10^{13}
M_\odot$.  The mass ratio and the ratio of the
Einstein radius change for the parameters $b$ and $r_{\rm s}$ in
similar fashion as for Abell 2163. The change of $b$ and $\theta_{\rm s}$
for Abell 2218 shows the same behavior in Table 2(a) for both $\nu=1$
and $\lambda=1$ groups. The reduced
$\chi^2$ values are smaller than the values for Abell 2163 because of
the smoothness of the X-ray emission. As seen in the mass estimate for
Abell 2163, both the isothermal sphere and the best fit-$\beta$-model
mass models cannot produce giant arcs for Abell 2218. The cusped
density models, however, produce finite Einstein radii,
however these are
less than the estimated radius. For cusped density models, the ratio,
$m_{\rm lens}/m_{\rm X}$ and $\theta_{\rm E,X}/\theta_{\rm E}$
approach one as $\mu$ increases. For Abell 2218, our best model is
$(\mu,\nu,\lambda)$=(1.4,1,1.6). The estimated Einstein radius for our
best model for Abell 2218 corresponds to 16\% of the observed
value. The projected mass $m_{\rm X}$ for our best model is 27\%
larger than the $\beta$-model mass. We cannot reduce the mass
discrepancy of Abell 2218 as significantly as we did for Abell
2163. This is mainly due to the different gas temperature
of the two clusters.

\subsection{RX J1347.5-1145}

A distant ($z_{\rm l}=0.451$) cluster RX J1347.5-1145 has two dominant
galaxies. It is the most X-ray luminous cluster: $L_{\rm X}=7.3 \times
10^{45}$ ergs ${\rm s^{-1}}$ (0.1-2.4 keV) and $T$=9.3\,keV(Schindler
et al. 1997). The X-ray profile of this cluster has a strong peak.
Schindler et al. (1997) noted that this strong peak is probably
evidence of cooling flows. However, they could not find significant
evidence for cool gas in the spectra. Two bright arcs are located at
about $35''$ from the cluster center and are positioned on opposite
sides of the central galaxy (Schindler et al. 1995). The redshift of
these arcs is 0.81 (Sahu et al. 1998).

In Figure 5, we present the observed surface profile of RX
J1347.5-1145 derived from the {\it ROSAT} HRI images (Schindler et al. 
1997) and the best-fit profiles estimated from various models. As a
result of fitting by the $\beta$-model, we find an extremely small
core radius $r_{\rm c}$ of 56kpc in comparison with that of nearby
clusters, $r_{\rm c}\approx 250$kpc (Jones \& Forman 1984). Indeed,
the core radii of Abell 2163 and Abell 2218 yield about 200-300kpc in
our fitting. The observed profile is steeper than the the best-fit
$\beta$-model inside the core radius. This steep profile with a small
core radius is presumably evidence of cooling flows, as Schindler et
al. (1997) suggested. As shown in Figure 5, all of the best-fit
profiles computed from the cusped density profiles is steeper than the
best-fit $\beta$-model. The surface brightness for $\mu \leq 1.6$ of
both $\nu=1$ and $\lambda$=1 group are in excellent agreement with the
data, because of the steepness of the data points at the center of
this cluster.

The projected lensing mass of RX J1347.5-1145 is $6.26\times10^{14}
M_\odot$. The uncertainty of the mass is caused by the unknown
redshift of the lensed source galaxy.  Table 2(c) indicates that the
reduced $\chi^2$ values in RX J1347.5-1145 are approximately equal to
one for both the $\mu=1$ and $\nu=1$ groups, except for the models
with $\mu=1.8$. Thus, models with $\mu \leq 1.6$ are
acceptable. However, the mass discrepancy is not decreased for RX
J1347.5-1145. The estimated Einstein radius and the projected X-ray
mass are almost unchanged, even if $\mu$ increases. The projected mass
$m_{\rm X}$ for $(\mu,\nu,\lambda)$=(1.6,1,1.4) model is only 7\%
larger than the $\beta$-model. This may be due to the small core
radius of this cluster. The X-ray core radius $\theta_{\rm c}$ of this
cluster is smaller than the Einstein radius $\theta_{\rm E}$, although
the X-ray core radius is large than the Einstein radius for the
previous two clusters. As shown in Figure 5, the surface brightness
profiles are almost unchanged at $\theta>\theta_{\rm c}$, even if the
model is changed. Therefore, we conclude that the projected mass
discrepancy of RX J1347.5-1145 was not decreased by our method based
on available {\it ROSAT} data. The {\it ROSAT} HRI data may be
dominated by X-rays from the cooling flow, not from the cluster as a
whole, however. To improve our analysis for RX J1347.5-114, we need the X-ray
data from the cluster as a whole, and we should take cooling flows
into account.

\section{Conclusions and Discussion}

\indent

We studied the effect of the density cusp on the mass cluster
estimates. We compared the Einstein radii estimated from the density
profile with a cusp with the observed radius. We showed that the
marginally steep cusp, $\mu=1.4$, reduces the discrepancy of the mass
estimates for Abell 2163 and Abell 2218. The ratio of the lensing mass
to the X-ray mass is improved from 2.2 to 1.4 for Abell 2163, and from
3 to 2.4 for Abell 2218. 

We found that the clusters should have a density cusp at the center to
reduce the discrepancy between the X-ray mass and the lensing
mass. The cusped density model is consistent with the density profile
found in the N-body simulations (Navarro et al. 1997). Navarro et
al. (1997) concluded that the density profile behaves as $\sim r^{-1}$
toward the center. However, the more strongly cusped density profile
at the center is expected to reconcile the mass discrepancy. In fact,
our best model falls off $\rho\propto r^{-1.4}$ around the center. The
density profiles of clusters of galaxies might be more strongly peaked
at the center as Fukushige \& Makino (1997) and Moore et al. (1997)
have shown with N-body simulations.

Although the introduction of a cusp in the density profile is not
sufficient to resolve the mass discrepancy, the density cusp mainly
contributes to the reduction of the discrepancy. In addition to
density cusps, we should take into account the gas temperature
decline, the second mass clump and the gas dynamical motion.
We expect that ellipticity in the mass distribution plays an important
role in resolving the discrepancy. Indeed, Bartelmann (1996) has shown
that mass estimation under the assumption of a spherically symmetric
mass distribution are likely to overestimate the mass by $\sim$
30\%-50\% . If the density cusp model reduces the ratio of the lensing
mass to the X-ray mass $m_{\rm lens}/ m_{\rm X}$ to $1.3-1.6$, the
mass discrepancy would be resolved by the adoption of a more detailed
cluster model, especially elliptical mass distributions for non-cooling
flow clusters (Asano \& Makino 1998).
For RX J1347.5-1145, the difference between the lensing mass and the
X-ray mass was not improved. This may be due to cooling flows
(Schindler et al. 1997). For this cluster the cooling flows should be
taken into account (Allen 1997).

Furthermore, independent of X-ray observations it has been recognized
that the cusped models with $1 \leq \mu \leq 1.4$ are more plausible
in gravitational lensing analysis. Bartelmann (1996) pointed out that
the NFW profile predicts large values of the radial magnification for
the cluster of galaxies MS 2137. This problem implies that the
corresponding sources must be surprisingly thin in the radial
direction. Evans \& Wilkinson (1997) have shown that lenses which have
more singular density cusps than the NFW profile assuage Bartelmann's
problem. The result of Evans \& Wilkinson, which was derived from the
analysis of gravitational lensing, agrees well with our result.

We adopted a density model which declines as $\sim r^{-3}$ at large
radii. However, the density profile at large radii is uncertain. In
N-body simulations it is difficult to determine whether mass clumps at
large radii are assigned to the cluster. 
In fact, the profile of the
Hernquist model for elliptical galaxies decreases as $\sim r^{-4}$ at large
radii. Dubinski \& Carlberg (1991) have shown by numerical simulations
that the profile of clustering objects falls off as $\sim r^{-4}$ at
large radii. Furthermore, Markevitch \& Vikhlinin (1997) have applied the
mass density model $\rho\propto x^{-1}(1+x)^{1-\alpha}$ to Abell
2256, and fit the data with the surface brightness
estimated from mass models for $\alpha<5$. However, mass densities
which decline faster than  $r^{-3}$ are less effective for the
formation of giant arcs than our model.

Miralda-Escud\'{e} \& Babul (1995) have studied how cusped density
models resolve the mass discrepancy. They have shown that a cusped
mass density produces a peak in the isothermal gas density around
the center, and that the cusped density model is inconsistent with
{\it ROSAT} PSPC data of Abell 2218, Abell 2163 and Abell 1689. The
{\it ROSAT} PSPC data show extremely small errors. On the other hand,
we have used the {\it ROSAT} HRI data, which have relatively large
errors. The two detectors are very different in the spatial resolution
and their backgrounds. In addition, the fit is
sensitive to the existence of substructures in the clusters, and the
results depend on the data. If the surface brightness estimated from
our models is fit to the {\it ROSAT} PSPC data, the mass
discrepancy would not be significantly reduced.

Lastly, we comment on the scaling parameter $r_{\rm s}$ in the NFW
profile. By numerical simulations of dark halos, Navarro, Frenk, \&
White (1996) predicted that the scaling parameter $r_{\rm s}$ is
approximately several hundred kpc at the cluster mass scale.  On the
other hand, $r_{\rm s}$ derived from our analysis are about 1Mpc for
the Abell 2163 and 2218 clusters. Namely, the size of $r_{\rm c}$
predicted by Navarro, Frenk, \& White is smaller than the observed
values. This problem has been already pointed out by Makino et
al. (1997). Both Bartelmann (1997) and Evans \& Wilkinson (1997)
adopted the scaling parameter which satisfies the estimation by the
numerical simulations of Navarro, Frenk, \& White. Their results for
MS 2137 will be altered by taking into account the constraint by X-ray
observations.

\vspace{1.5cm}

We are grateful to Makoto Hattori, Yasushi Suto and Kenji Tomita for
their helpful advices, and Doris Neumann for providing the X-ray
surface brightness data of Abell 2163 and Abell 2218. We thank Sabine
Schindler for kindly providing the {\it ROSAT} HRI data of RX
J1347.5-1145 in ASCII format. Lastly, we appreciate Maxim Markevitch
for kindly providing the {\it ROSAT} HRI images for Abell 2218.

\newpage

\begin{center}
{\bf \LARGE References}
\end{center}

\begin{description}
\item
Allen, S. W. 1998,  MNRAS 296, 392
\item
Allen, S. W., Fabian, A. C., \& Kneib, J. P. 1996, MNRAS 279, 615
\item
Asano, K., \& Makino, N. 1998 in preparation.
\item
Bartelmann 1996, A\&A 313, 697
\item
Birkinshaw, M., \& Hughes, J. P. 1994, ApJ 420, 33
\item
Carlberg, R. G., Yee, H. K. C., Ellingson, E., Morris, S. L., Abraham. 
R., Gravel, P., Pritchet, C. J., Smecker-Hane, T., Hartwick, F. D. A.,
Hesser, J. E., Hutchings, J. B., \& Oke, J. B. 1997, ApJ 485, L13
\item
Dubinski, J., \& Carlberg, R. 1991, ApJ 378, 496
\item
Eke, V. R., Navarro, J. F., \& Frenk, C. S. 1997, to appear in ApJ
(astro-ph/9708070)
\item
Elbaz, D., Arnaud, M., B\"{o}hringer, H. 1995 A\&A 293, 337
\item
Evans, N. W., \& Collett, J. L. 1997, ApJ 480, L103
\item
Evans, N. W., \& Wilkinson, M. 1998, MNRAS 296, 800
\item
Evrard, A. E., Metzler, C. A., \& Navarro, J. F. 1996, ApJ 469, 494
\item
Fukushige, T., \& Makino, J. 1997, ApJ 477, L9
\item
Grossman, S. A., Narayan, R. 1989, ApJ 344, 637
\item 
Henry, J. P., Briel, U. G.\& Nulsen, P. E. J. 1993, A\&A 271, 413
\item
Hernquist, L. 1990, ApJ 356, 359
\item
Holzapfel, W. L., Arnaud, M., Ade, P. A. R., Church, S. E.,
Fischer, M. L., Mauskopf, P. D., Rephaeli, Y., Wilbanks, T. M.,
\& Lange, A. E. 1997, ApJ 480, 449
\item
Jones, C., \& Forman, W. 1984, ApJ 276, 38
\item
Kneib, J. P., Mellier, Y., Pell\'o, R., Miralda-Escud\'e, J.,
Le Borgne, J. -F., B\"ohringer, H.,
\& Picat, J. -P.
1995, A\&A 303, 27
\item
Le Borgne, J. F., Pello, R., \& Sanahuja, B. 1992, A\&AS 95, 87
\item
Loeb, A., \& Mao, S. 1994, ApJ 435, L109
\item
Makino, N. 1996, PASJ 48, 573
\item
Makino, N. 1997, ApJ 490, 641
\item
Makino, N., Sasaki, S., \& Suto, Y. 1998, ApJ 497, 555
\item
Maoz, D., Rix, H-W., Gal-Yum, A., Gould, A. 1997, ApJ 486, 75
\item
Markevitch, M. 1997, ApJ 483, L17
\item
Markevitch, M., \& Vikhlinin, A. 1997, ApJ 491, 467
\item
Miralda-Escud\'{e}, J. 1993, ApJ 403, 497
\item
Miralda-Escud\'e, J., \& Babul, A. 1995, ApJ 449, 18
\item
Moore, B., Governato, F., Quinn, T., Stadel, J., \& Lake, G. 1998,
ApJ, 499, L5
\item
Mushotzky, R. F., \& Loewenstein, M. 1997, ApJ 481, L63
\item
Navarro, J. F., Frenk, C. S., \& White, S. D. M. 1995, MNRAS 275, 720
\item
Navarro, J. F., Frenk, C. S., \& White, S. D. M. 1996, ApJ 462, 563
\item
Navarro, J. F., Frenk, C. S., \& White, S. D. M. 1997, ApJ 490, 493
\item
Pell\'{o}, R., Le Borgne, J. F., Sanahuja, B., Methez, G., Fort, B. 
1992, A\&A 266, 6
\item
Sarazin, C. 1988, X-ray Emissions from Clusters of Galaxies.
Cambridge Univ. Press, Cambridge
\item
Sahu, K. C., Shaw, R. A., Kaiser, M. E., Baum, S. A., Ferguson, H. C.,
Hayes, J. J. E., Gull, T. R., Hill, R. J., Hutchings, J. B.,
Kimble, R. A., Plait, P., \& Woodgate, B. E.
1998, ApJ 492, L125
\item
Schindler, S., Guzzo, L., Ebeling, H., B\"ohringer, H., Chincarini, G.,
Collins, C. A., De Grandi, S., Neumann, D. M., Briel, U. G.,
Shaver, P., \& Vettolani, G.
1995, A\&A 299, L9
\item
Schindler, S., Hattori, M., Neumann, D. M., \& B\"ohringer, H.
1997, A\&A 317, 646
\item
Squires, G., Kaiser, N., Babul, A., Fahlman, G., Woods, D.,
Neumann, D. M., \& B\"{o}hringer, H. 
1996, ApJ, 461, 572
\item
Suto, Y., Sasaki, S., \& Makino, N. 1998 in preparation.
\item
Wu, X. -P., \& Fang, L. -Z. 1996, ApJ 461, L5
\end{description}
\newpage

\begin{center}
{\bf \large Figure captions}
\end{center}

\figcaption{ The X-ray surface brightness profiles estimated from the
mass density models. The parameter $b$ is equal to 0.7.  (a)the
surface brightness profiles for the $\nu=1$ group.  The solid line
indicates the surface brightness for the models with the parameters
$(\mu,\nu,\lambda)=(1,1,2)$. The dotted line, the short dashed
line, the long dashed line, and the dotted-dashed line show the
surface brightness for the models $(\mu,\nu,\lambda)=(1.2,1,1.8)$,
(1.4,1,1.6), (1.6,1,1.4), (1.8,1,1.2), respectively.  (b) the
surface brightness profiles for $\lambda=1$ group. The solid line
indicates the surface brightness for the models with the parameters
$(\mu,\nu,\lambda)=(1,2,1)$. The dotted line, the short dashed
line, the long dashed line, and the dotted-dashed line show the
surface brightness for the models $(\mu,\nu,\lambda)$=(1.2,1.8,1),
(1.4,1.6,1), (1.6,1.4,1), (1.8,1.2,1), respectively.}

\figcaption{The X-ray surface brightness profile of the X-ray emission
from Abell 2163. The boxes are the data of the observed surface
brightness. The {\it ROSAT} HRI data were kindly provided to us by Doris
Neumann. The solid lines are the best-fit $\beta$-model with
parameters, $\beta=0.54$ and $\theta_{\rm c}=65''$.
(a) We present the surface brightness for the isothermal sphere with a
finite core radius model and the $\nu=1$ group for $\mu \leq 1.2$. The
dotted line indicates the surface brightness for the isothermal sphere
$(\mu,\nu,\lambda)=(0,2,1)$. The short dashed line and the long
dashed line show the surface brightness for the cusped models with
parameters $(\mu,\nu,\lambda)=(1,1,2)$ and (1.2,1,1.8),
respectively.
(b) The X-ray surface brightness for the $\nu=1$ groups
with $\mu \geq 1.2$.  The dotted line, short dashed line and the long
dashed line show the surface brightness for the cusped models with
parameters $(\mu,\nu,\lambda)=(1.4,1,1.6)$, (1.6,1,1.4), and
(1.8,1,1.2), respectively.
(c) The X-ray surface brightness for the
$\lambda=1$ groups with $\mu \leq 1.4$.  The dotted line, short dashed
line and the long dashed line show the surface brightness for the
cusped models with parameters $(\mu,\nu,\lambda)=(1,2,1)$,
(1.2,1.8,1), and (1.4,1,6.1), respectively.}

\figcaption{The reduced $\chi^2$ values, the ratios of the X-ray
estimated Einstein radius to the observed Einstein radius and the
ratios of the lensing mass to the X-ray mass for Abell 2163 and Abell
2218.  We present the results only for the $\nu=1$ group. The 
behavior for the $\lambda=1$ group is similar to the $\nu=1$ group.
The solid lines and dotted lines indicate the quantities of Abell
2163 and Abell 2218, respectively.
(a) the reduced $\chi^2$ values for various values of the parameter $\mu$.
(b) the ratios of the X-ray estimated Einstein radius to the observed Einstein 
radius. 
(c)the ratios of the lensing mass to the X-ray mass
}

\figcaption{The X-ray surface brightness for Abell 2218. The {\it
ROSAT} HRI data were kindly provided to us by Doris Neumann. The
parameters for the best-fit $\beta$-model are $\beta=0.56$ and
$\theta_{\rm c}=53''$. We show the parameters for the other models in
Table 2(b).
(a) The surface brightness for the isothermal sphere with a
finite core radius model and the $\nu=1$ group for $\mu \leq 1.4$. The
dotted line indicates the surface brightness for the isothermal
sphere. The short dashed line, the long dashed line and the short and
dashed line indicate the surface brightness profiles for the cusped models with
parameters $(\mu,\nu,\lambda)=(1,1,2)$, (1.2,1,1.8), and (1.4,1,1.6)
respectively.
(b) The X-ray surface brightness for the
$\lambda=1$ groups with $\mu \leq 1.6$. The dotted line, short dashed
line, the long dashed line and the short and long dashed line show the
surface brightness for the cusped models with parameters
$(\mu,\nu,\lambda)=(1,2,1)$, (1.2,1.8,1), (1.4,1,6.1) and (1.6,1.4,1),
respectively.}

\figcaption{The X-ray surface brightness for RX J1347.5-1145. The {\it
ROSAT} HRI data were kindly provided to us by Sabine Schindler. The
parameters for the best-fit $\beta$-model are $\beta=0.56$ and
$\theta_{\rm c}=8.2''$.  In Table 2(c), we show the parameters for the
other models.
(a) The surface brightness for the isothermal sphere with a finite
core radius model and the $\nu=1$ group for $\mu \leq 1.4$. The dotted
line indicates the surface brightness for the isothermal sphere. The
short dashed line, the long dashed line and the short and dashed line
indicate the surface brightness profiles for the cusped models with
parameters $(\mu,\nu,\lambda)=(1,1,2)$, (1.2,1,1.8), and (1.4,1,1.6)
respectively.
(b) The X-ray surface brightness in RX J1347.5-1145 for the
$\lambda=1$ groups with $\mu \leq 1.6$. The dotted line, short dashed
line, the long dashed line and the short and long dashed line show the
surface brightness for the cusped models with parameters
$(\mu,\nu,\lambda)=(1,2,1)$, (1.2,1.8,1), (1.4,1,6.1) and (1.6,1.4,1),
respectively.}

\newpage

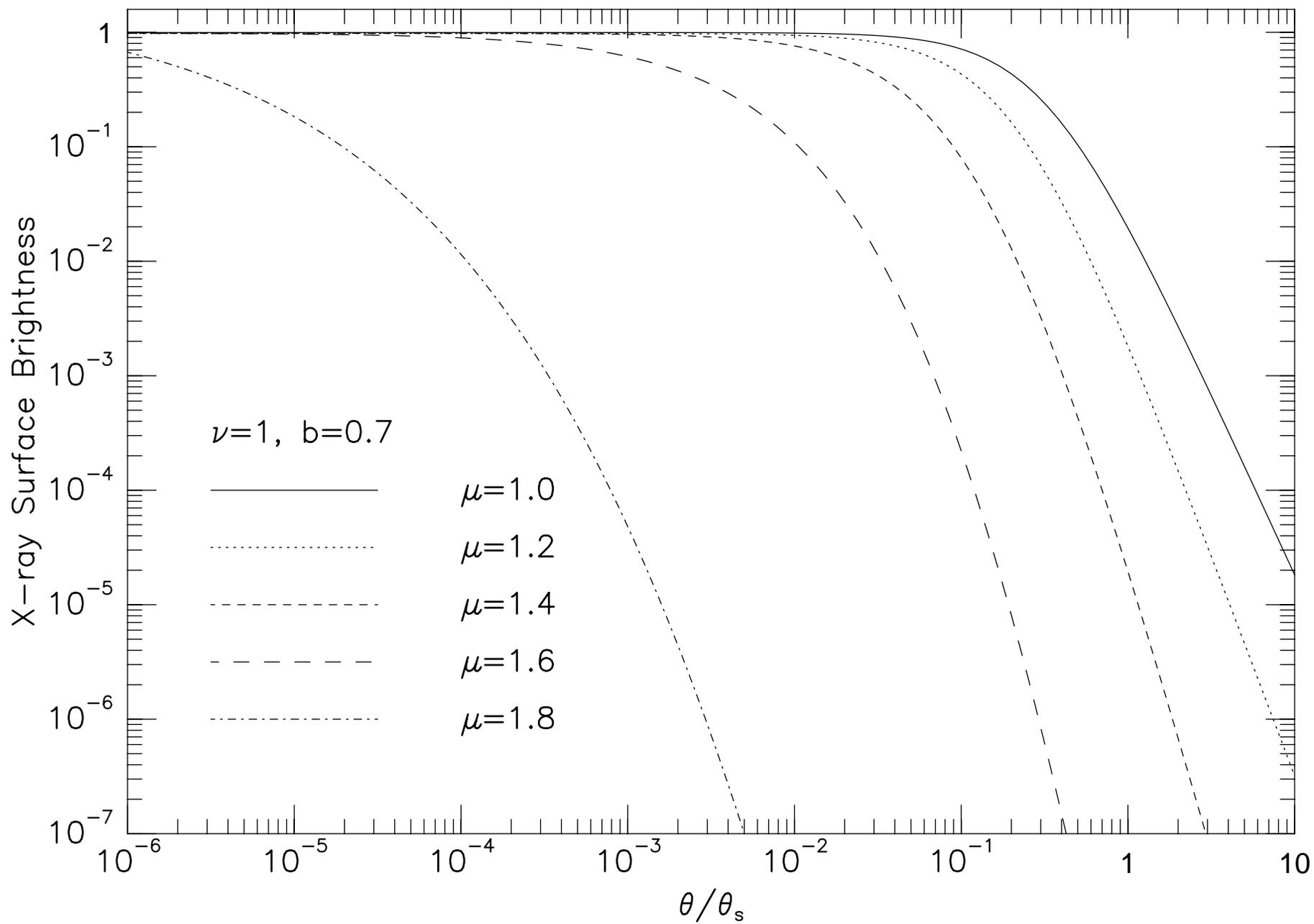
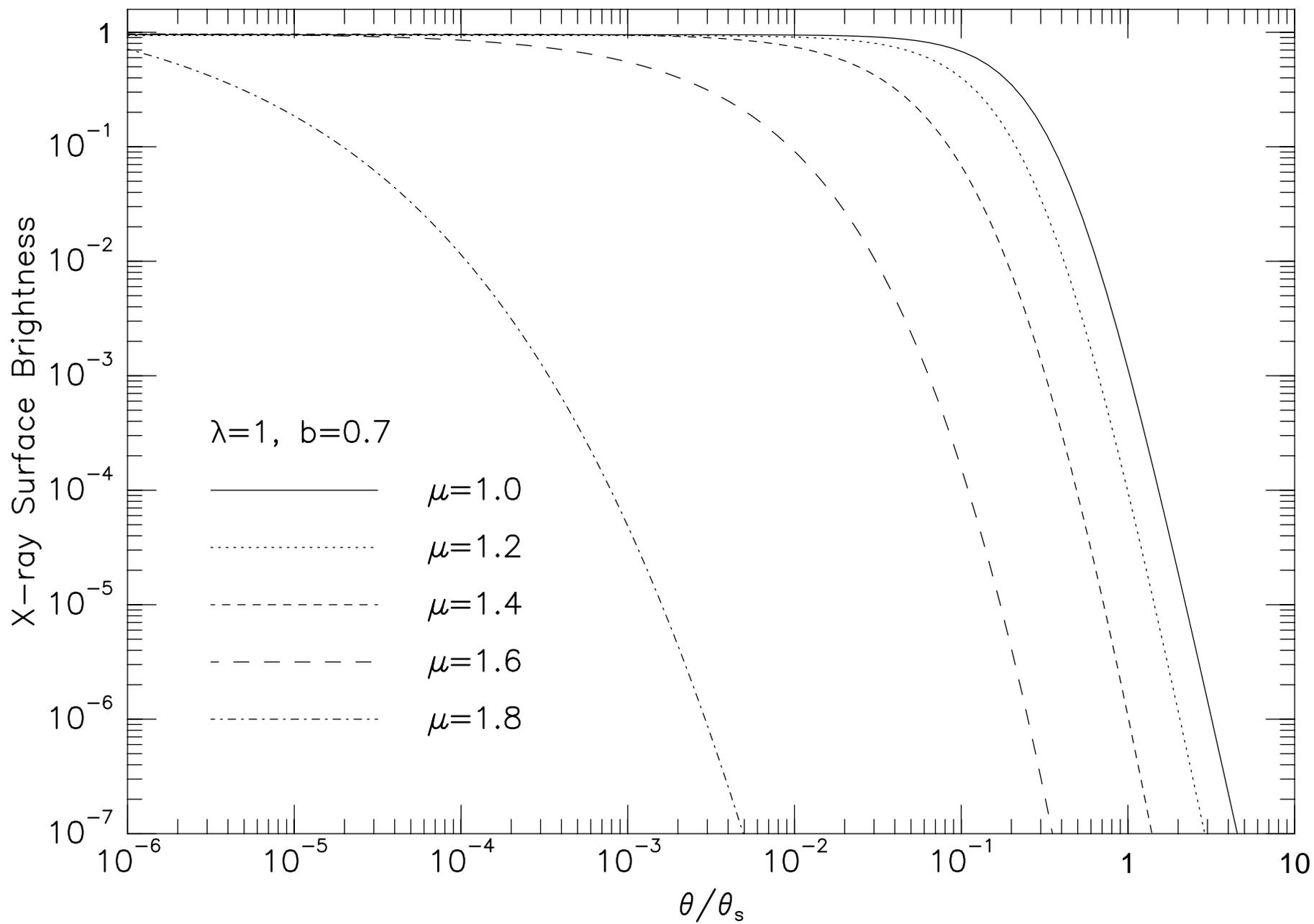
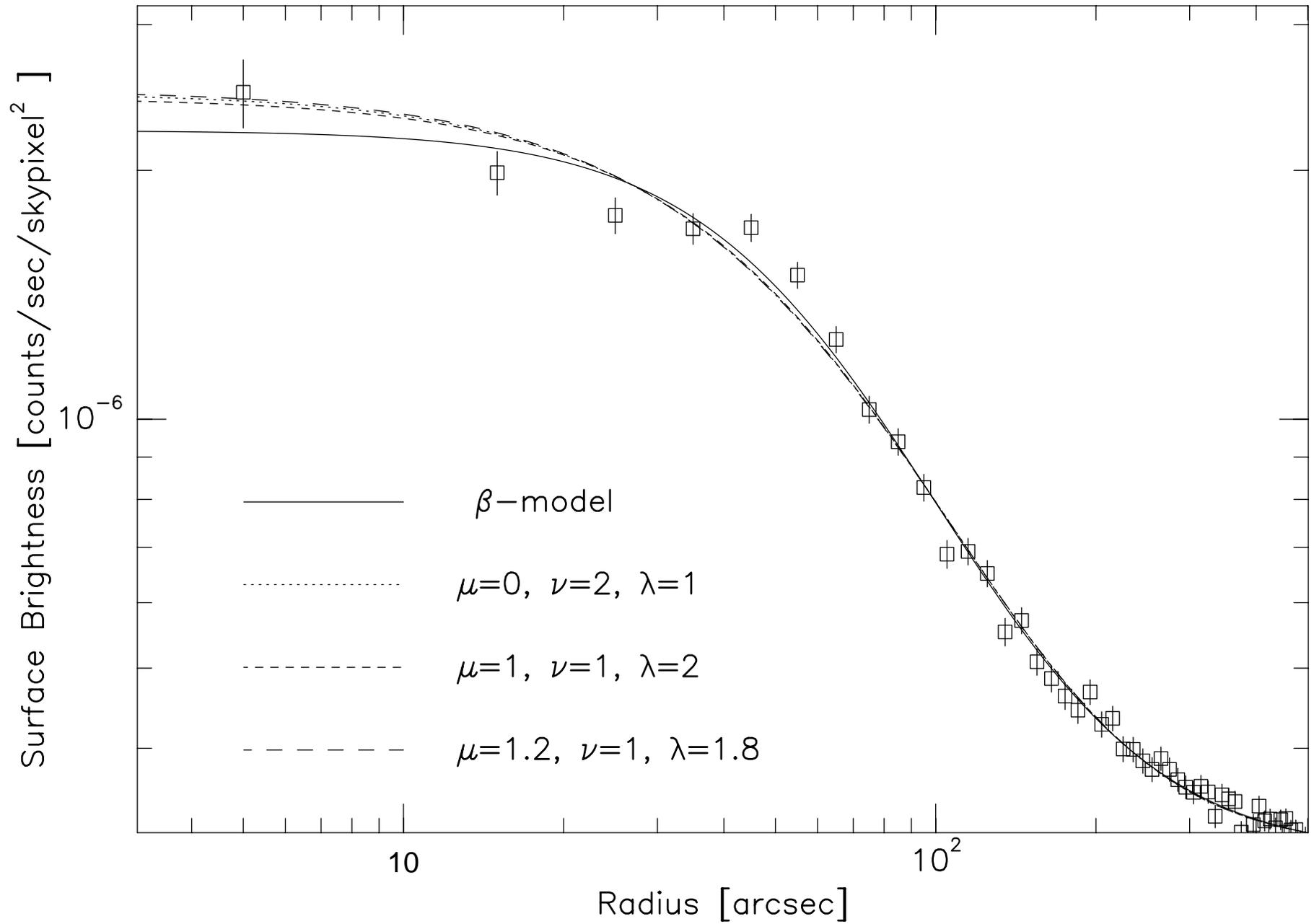
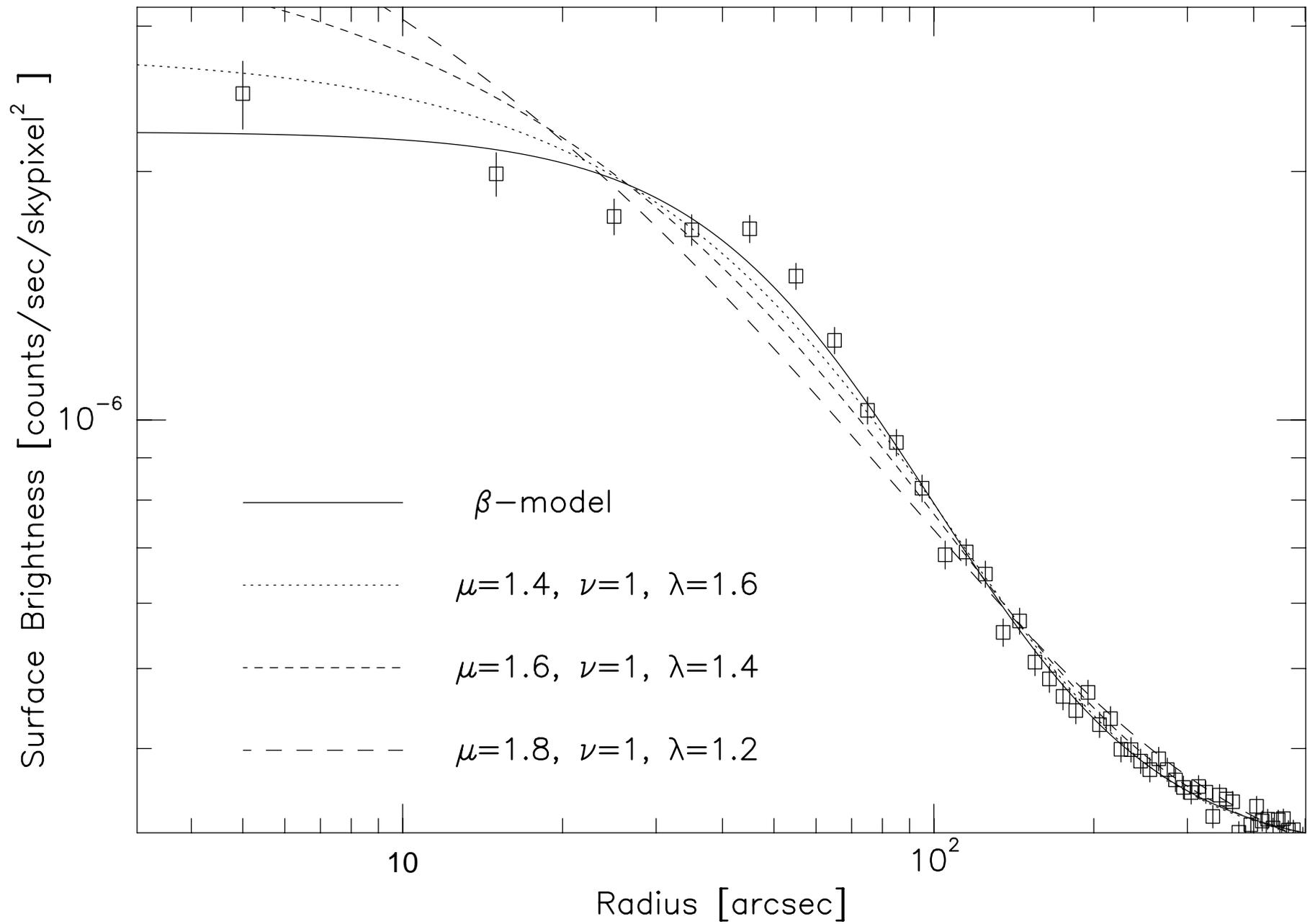
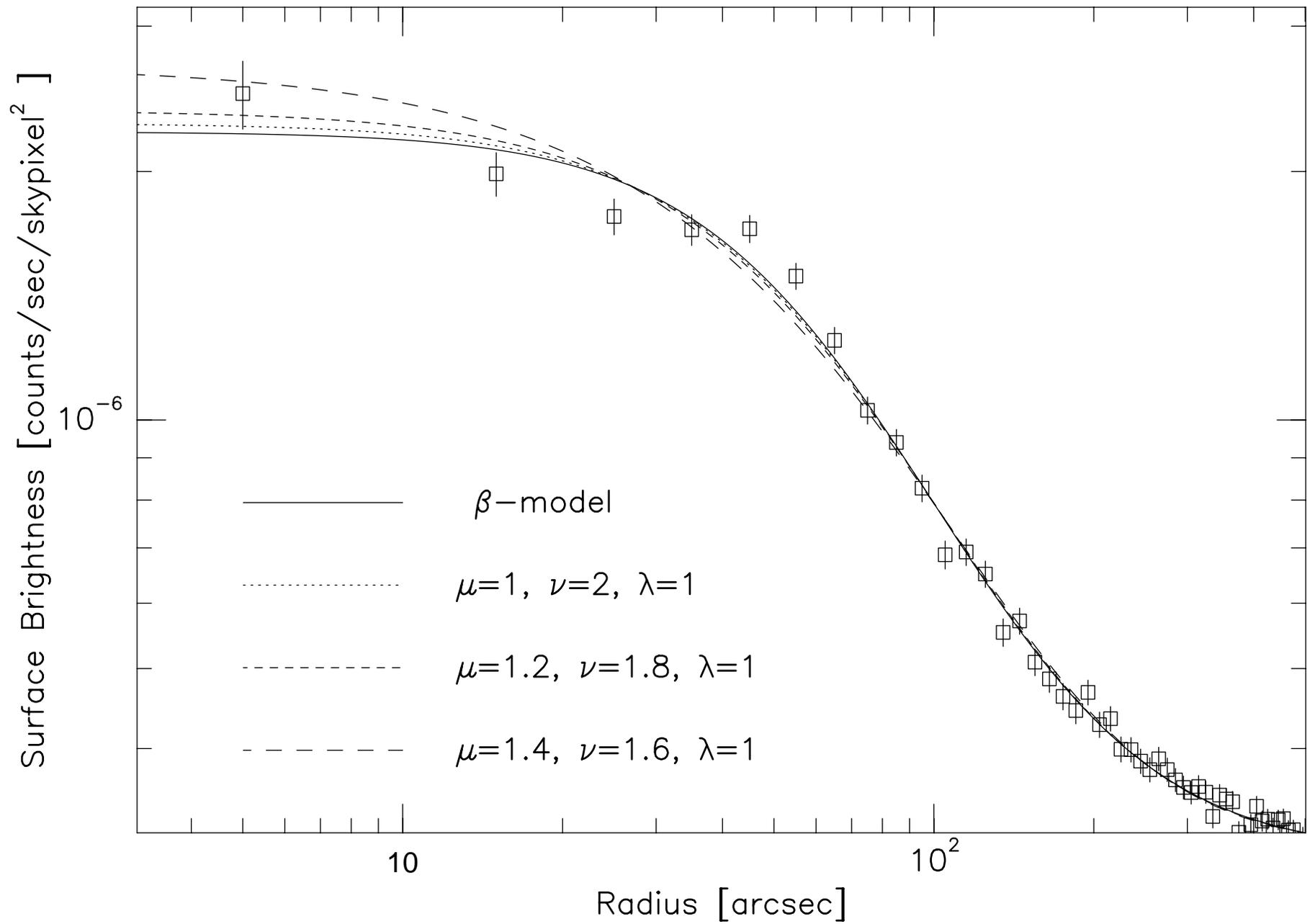
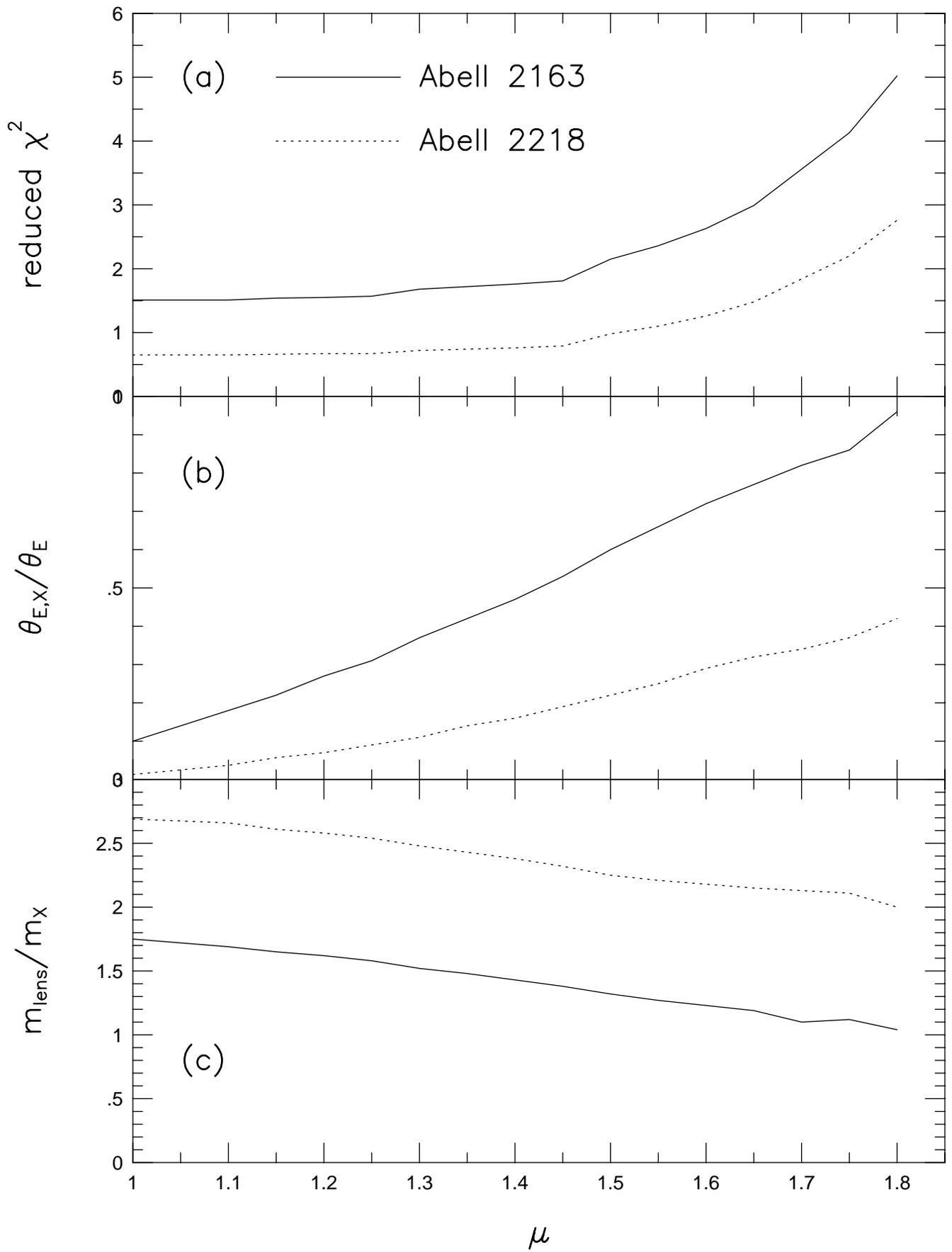
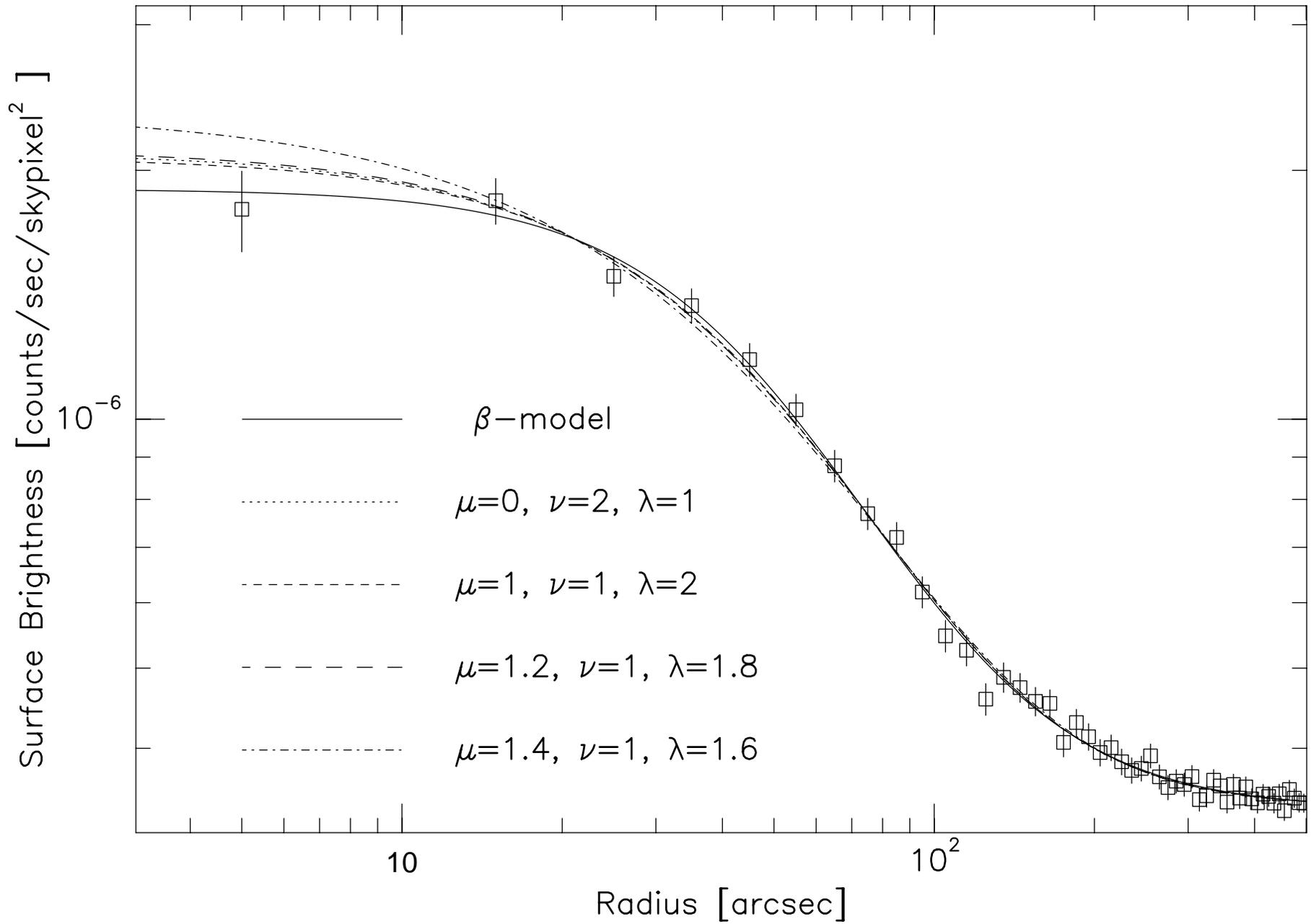
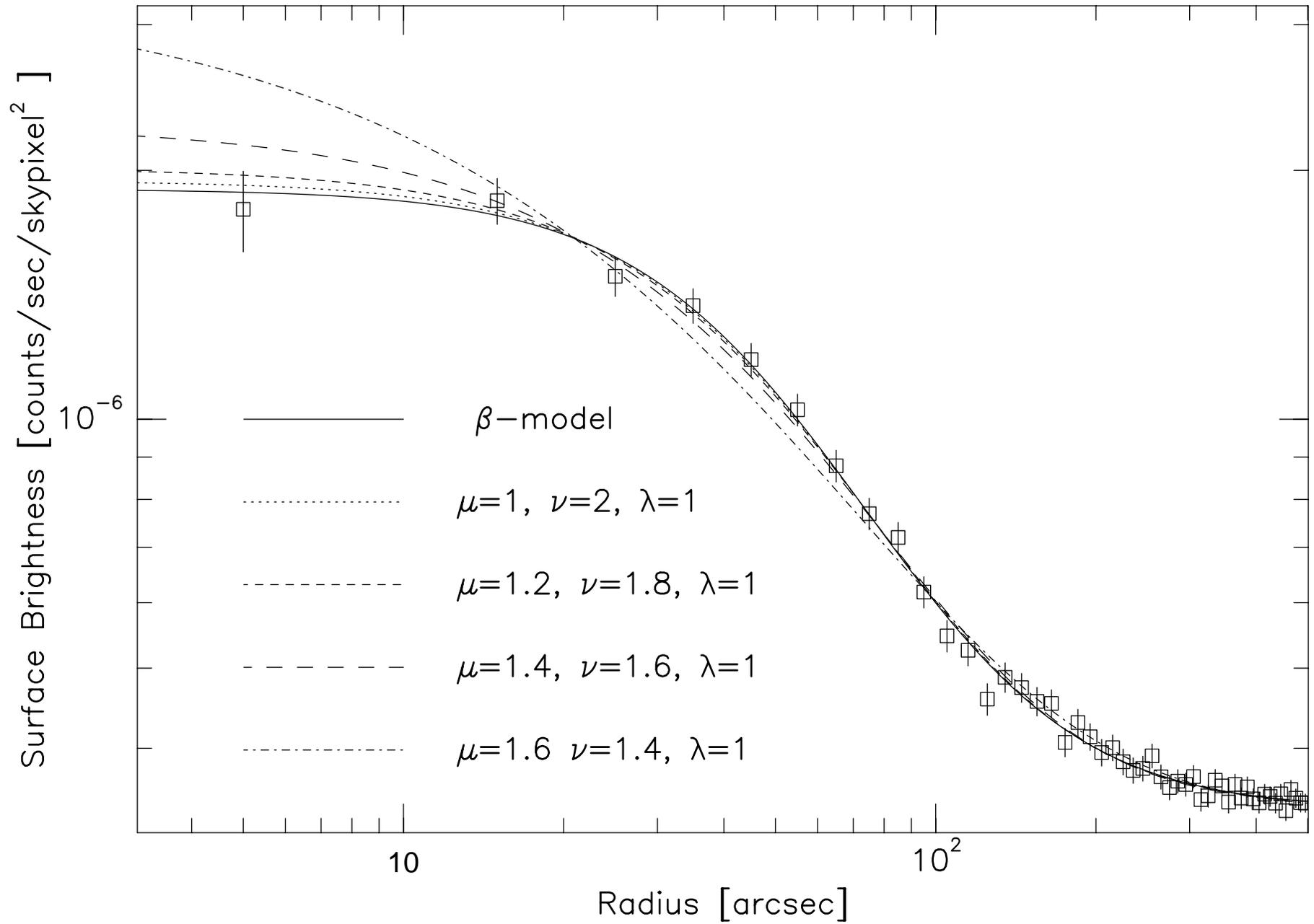
\begin{center}
\begin{tabular}{lcccccc}
\hline \hline
Cluster & $z_{\rm l}$ & $z_{\rm s}$ & $T$ 
& $\theta_{\rm E}$ & $\theta_{\rm c}$ &
$\beta$ \\
&&&keV&arcsec&arcsec & \\ \hline
Abell 2163 & 0.201 & 0.728 & 12.4 & 15.6 & 65 & 0.54\\
Abell 2218 & 0.175 & 0.702 & 7.2 & 20.8 & 53 & 0.56 \\
RX J1347.5-1145 & 0.451 & 0.81 & 9.3 & 35 & 8.2 & 0.56\\ \hline
\end{tabular}
\end{center}
{\footnotesize Table~1. Cluster sample. 
The parameters of $\beta$ and $\theta_{\rm c}$ are derived from
our analysis.

\begin{center}
\begin{tabular}{cccccc}
\hline \hline
$(\mu,\nu,\lambda)$ & $b$ & $\theta_{\rm c}/\theta_{\rm s}$ & $\theta_{\rm E,X}/
\theta_{\rm E}$ & $m_{\rm lens}/m_{\rm X}$ & reduced $\chi^2$ \\ \hline
$\beta$-model mass & - & - & - & 2.24 & 1.32 \\
$(0, 2, 1)$ & 0.15 & 1.9 & - & 2.01 & 1.55 \\
$(1, 1, 2)$ & 0.63 & 0.25 & 0.10 & 1.75 & 1.51 \\
$(1.2, 1, 1.8)$ & 0.56 & 0.15 & 0.27 & 1.62 & 1.55 \\
$(1.4, 1, 1.6)$ & 0.54 & 0.064 & 0.47 & 1.43 & 1.76 \\
$(1.6, 1, 1.4)$ & 0.49 & 0.015 & 0.72 & 1.23 & 2.63 \\
$(1.8, 1, 1.2)$ & 0.39 & 0.00043 & 0.96 & 1.04 & 5.02 \\
$(1, 2, 1)$ & 0.31 & 0.37 & 0.033 & 1.95 & 1.37 \\
$(1.2, 1.8, 1)$ & 0.32 & 0.25 & 0.18 & 1.75 & 1.44 \\
$(1.4, 1.6, 1)$ & 0.35 & 0.12 & 0.42 & 1.50 & 1.69 \\
$(1.6, 1.4, 1)$ & 0.36 & 0.032 & 0.70 & 1.24 & 2.58 \\
$(1.8, 1.2, 1)$ & 0.34 & 0.00067 & 0.89 & 1.10 & 5.12 \\ \hline
\end{tabular}
\end{center}
{\footnotesize Table~2(a). Results for Abell 2163}
\begin{center}
\begin{tabular}{cccccc}
\hline \hline
$(\mu,\nu,\lambda)$ & $b$ & $\theta_{\rm c}/\theta_{\rm s}$ & $\theta_{\rm E,X}/
\theta_{\rm E}$ & $m_{\rm lens}/m_{\rm X}$ & reduced $\chi^2$ \\ \hline
$\beta$-model mass & - & - & - & 3.02 & 0.59 \\
$(0, 2, 1)$ & 0.15 & 1.8 & - & 2.78 & 0.66 \\
$(1, 1, 2)$ & 0.66 & 0.24 & 0.013 & 2.69 & 0.65 \\
$(1.2, 1, 1.8)$ & 0.58 & 0.15 & 0.070 & 2.58 & 0.67 \\
$(1.4, 1, 1.6)$ & 0.57 & 0.060 & 0.16 & 2.38 & 0.76 \\
$(1.6, 1, 1.4)$ & 0.52 & 0.014 & 0.29 & 2.18 & 1.26 \\
$(1.8, 1, 1.2)$ & 0.39 & 0.00047 & 0.42 & 2.00 & 2.76 \\
$(1, 2, 1)$ & 0.33 & 0.37 & 0.0025 & 2.93 & 0.61 \\
$(1.2, 1.8, 1)$ & 0.33 & 0.25 & 0.045 & 2.74 & 0.63 \\
$(1.4, 1.6, 1)$ & 0.37 & 0.11 & 0.14 & 2.47 & 0.73 \\
$(1.6, 1.4, 1)$ & 0.37 & 0.030 & 0.28 & 2.20 & 1.24 \\
$(1.8, 1.2, 1)$ & 0.35 & 0.00063 & 0.39 & 2.11 & 2.82 \\ \hline
\end{tabular}
\end{center}
{\footnotesize Table~2(b). Results for Abell 2218}
\begin{center}
\begin{tabular}{cccccc}
\hline \hline
$(\mu,\nu,\lambda)$ & $b$ & $\theta_{\rm c}/\theta_{\rm s}$ & $\theta_{\rm E,X}/
\theta_{\rm E}$ & $m_{\rm lens}/m_{\rm X}$ & reduced $\chi^2$ \\ \hline
$\beta$-model mass& - & - & 0.27 & 2.89 & 1.03 \\
$(0, 2, 1)$ & 0.14 & 2.3 & 0.27 & 2.82 & 1.00 \\
$(1, 1, 2)$ & 0.60 & 0.29 & 0.25 & 2.96 & 1.00 \\
$(1.2, 1, 1.8)$ & 0.53 & 0.19 & 0.24 & 2.89 & 1.00 \\
$(1.4, 1, 1.6)$ & 0.45 & 0.098 & 0.25 & 2.81 & 1.01  \\
$(1.6, 1, 1.4)$ & 0.41 & 0.028 & 0.26 & 2.70 & 1.09  \\
$(1.8, 1, 1.2)$ & 0.58 & 0.000089 & 0.25 & 3.00 & 1.66 \\
$(1, 2, 1)$ & 0.32 & 0.39 & 0.23 & 3.12 & 1.02 \\
$(1.2, 1.8, 1)$ & 0.31 & 0.27 & 0.24 & 2.98 & 1.01 \\
$(1.4, 1.6, 1)$ & 0.31 & 0.16 & 0.24 & 2.87 & 1.01 \\
$(1.6, 1.4, 1)$ & 0.30 & 0.056 & 0.25 & 2.77 & 1.08 \\
$(1.8, 1.2, 1)$ & 0.47 & 0.00031 & 0.26 & 2.93 & 1.65 \\ \hline
\end{tabular}
\end{center}
{\footnotesize Table~2(c). Results for RX J1347.5-1145}

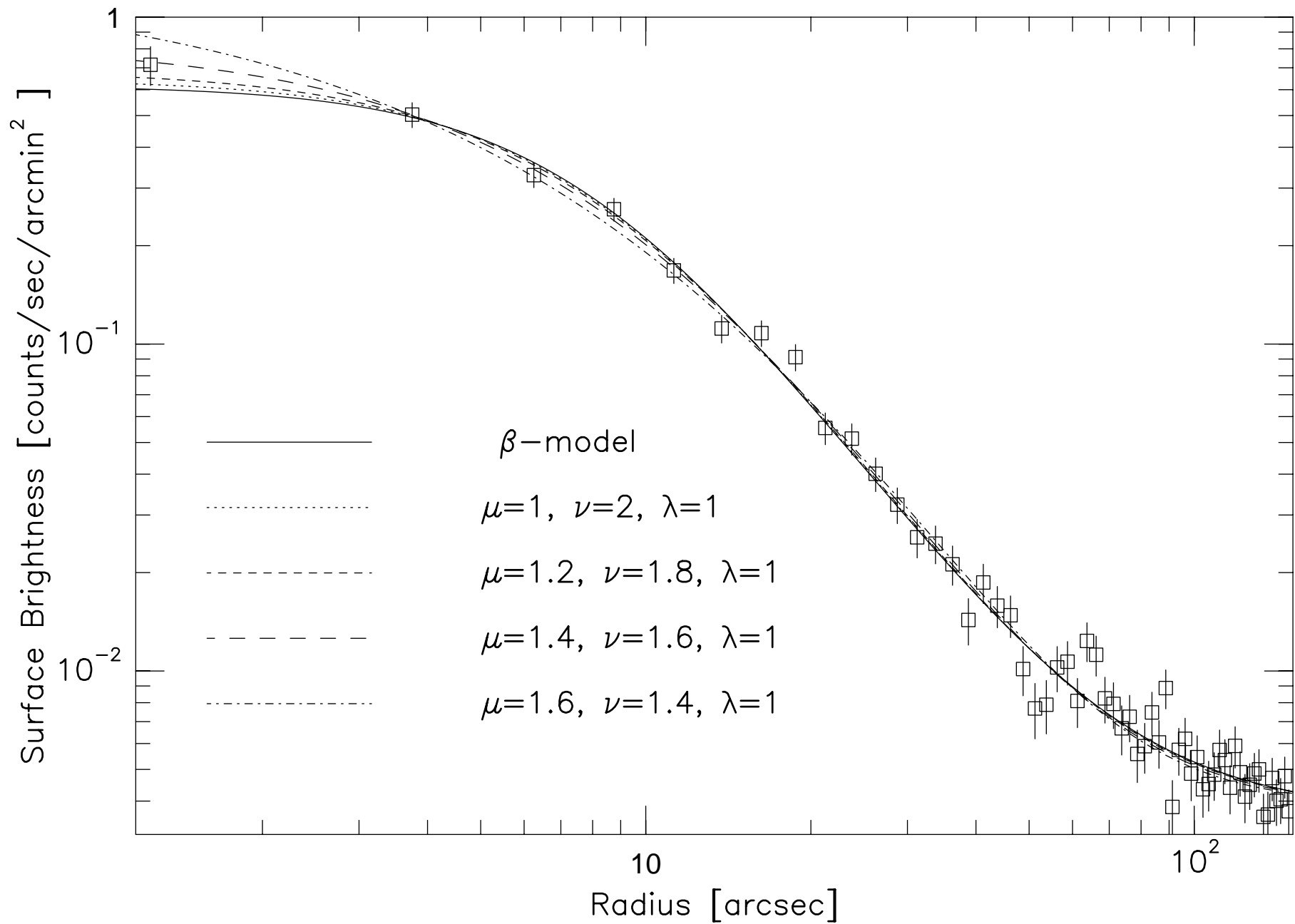
\end{document}